\documentstyle[12pt,epsfig]{article}

\newcommand{\f}{\frac}
\newcommand{\non}{\nonumber \\}
\newcommand {\beq}{\begin{equation}}
\newcommand {\eeq}{\end{equation}}
\newcommand {\beqa}{\begin{eqnarray}}
\newcommand {\beqal}{\begin{eqnarray}\label}
\newcommand {\eeqa}{\end{eqnarray}}
\newcommand {\bc}{\begin{center}}
\newcommand {\ec}{\end{center}}
\newcommand {\s}{\sigma}
\newcommand {\al}{\alpha^{'}}
\newcommand {\als}{\alpha^{'2}}
\newcommand {\th}{\theta}
\newcommand {\vth}{\vartheta}
\newcommand {\ka}{\kappa}
\newcommand {\pa}{\partial}
\newcommand {\de}{\delta}
\newcommand {\ep}{\epsilon}
\newcommand {\kpa}{k_{\parallel}}
\newcommand {\kpai}{k_{\parallel i}}
\newcommand {\kpaj}{k_{\parallel j}}
\newcommand {\kpe}{k_{\perp}}
\newcommand {\Tr}{\mbox{Tr}}

\def\vs5{\vspace*{5mm}}
\def\vs1{\vspace*{1cm}}
\def\vs2{\vspace*{2cm}}
\def\hs5{\vspace*{5mm}}
\def\hs1{\hspace*{1cm}}

\begin{document}

\title{
\hfill\parbox{4cm}{\normalsize IMSC/2005/03/01}\\
\vspace{1cm}
Aspects of Open-Closed Duality in a Background
$B$-Field
\author{S. Sarkar and B. Sathiapalan \footnote{email: 
\{swarnen, bala\}@imsc.res.in}\\
\\
{\em Institute of
Mathematical Sciences}\\
{\em Taramani}\\{\em Chennai, India 600113}}}
\maketitle

\begin{abstract}
\noindent
We study closed string exchanges in background $B$-field. 
By analysing the two point one loop amplitude in bosonic string theory, 
we show that tree-level exchange of lowest lying, tachyonic and massless 
closed string modes, have IR singularities similar to those of 
the nonplanar sector in noncommutative 
gauge theories. We further isolate the contributions from each of the 
massless modes. We interpret these results as the manifestation of 
open/closed string duality, where the IR behaviour of the boundary 
noncommutative gauge theory is reconstructed from the bulk theory of 
closed strings.    
\end{abstract}

\newpage

\section{Introduction}

Dynamics of strings in background fields is an old subject. Open 
string dynamics in background two form field is an 
interesting area giving rise to various new phenomena. Specifically the 
study of open strings in the presence of background constant 
two from $B$-field has shown how noncommutative spacetimes can arise in 
string theory \cite{douglas1,schomerus,chu,Seiberg}. By switching on a 
constant $B$-field along the 
world-volume directions of a $D$-brane, it was shown that spacetime 
coordinates along these directions no longer commute. The low energy 
dynamics of the $D$-brane is described by a noncommutative Yang-Mills 
theory. Noncommutative gauge theories and noncommutative versions of 
other field theories have since been studied extensively for the past 
few years. For reviews on the subject see \cite{douglas}. These field 
theories are a class of nonlocal theories, but 
are tractable and offer new interesting phenomena that are closely 
related to the parent string theory.

A generic characteristic of noncommutative field theories is the mixing 
of the UV and IR regimes arising in the nonplanar sectors 
\cite{minwalla}. A lot of effort by 
various authors have been made to understand this interesting feature. 
Usual 
notions of Wilsonian RG do not fit in the continuum limit. 
Contradictory, as this may seem in the field theoretic picture, this 
phenomenon has a natural interpretation in string theory. Open string 
one loop amplitudes have a dual description in terms of tree-level 
propagation of closed string modes. The UV region of the open string 
loop, in the dual picture is dominated by the lightest closed string 
modes. The UV divergences of the open string loop can thus be 
reinterpreted as IR divergences due to propagating massless closed string modes.
Though attempts have been made along these lines to understand the IR 
divergences occurring in noncommutative field theories by integrating out 
high energy degrees of freedom \cite{oneloop, rajaraman}, the 
picture 
still remains unclear. See also \cite{Chaudhuri}. We address 
this issue in this paper. In 
the bosonic string theory setting, we first analyse the two-point one 
loop amplitude for gauge bosons on the brane, in the closed string 
channel. We argue that the region of the modulus giving rise to 
divergences (that are regulated in the nonplanar amplitudes) in 
noncommutative field theories can be identified as the region where the 
lightest closed string modes dominate in the dual picture. In usual 
quantum field theory, because of infinities, there is no way to compare 
quantitatively the contributions in these two pictures. In the presence
of the background $B$-field the nonplanar diagrams are regulated and 
a quantitative approach can thus be made. The full two point 
open string amplitude also contains finite contributions which would 
require the entire tower of closed string states for its dual 
description. However, the singular IR behaviour of the nonplanar 
amplitudes, in the boundary noncommutative gauge theory can be seen from 
the exchange of closed strings in the bulk. On the broader side,  
these results can be seen from the point of view of 
open/closed string duality. This is similar in 
spirit to the AdS/CFT correspondence \cite{maldacena}. 
Here the bulk theory of closed strings is in flat space, 
but in the presence of constant background $B$-field.    

Though there are additional 
tachyonic divergences, we are able to show that the form of IR 
divergences with appropriate tensor structures can be extracted by 
considering only lowest lying modes (tachyonic and massless). We further 
analyse the two point amplitude by studying massless closed string 
exchanges in background constant $B$-field. From this analysis we are 
able to isolate the individual contributions from the massless closed 
string exchanges.

This paper is organised as follows. In section 2, we review concisely, 
the open string dynamics in the presence of background constant $B$-field 
and the arising of noncommutative field theory in the Seiberg-Witten 
limit. In section 3, we study the one loop open string amplitude 
in the UV limit and write down the contribution from the lowest  
states. In section 4, we analyse massless closed string exchange in 
background $B$-field and reconstruct the massless contribution computed 
in
section 3. In section 5, we conclude with discussions and further 
prospects.

\noindent
{\it Conventions:} We will use capital letters $(M, N,...)$ to denote 
general spacetime indices and small letters $(i,j, ...)$ for coordinates
along the $D$-brane. Small Greek letters $(\alpha,\beta...)$ will be 
used 
to denote indices for directions transverse to the brane.

\def\limits{\noindent

\begin{figure}[t]
\begin{center}
\epsfig{file=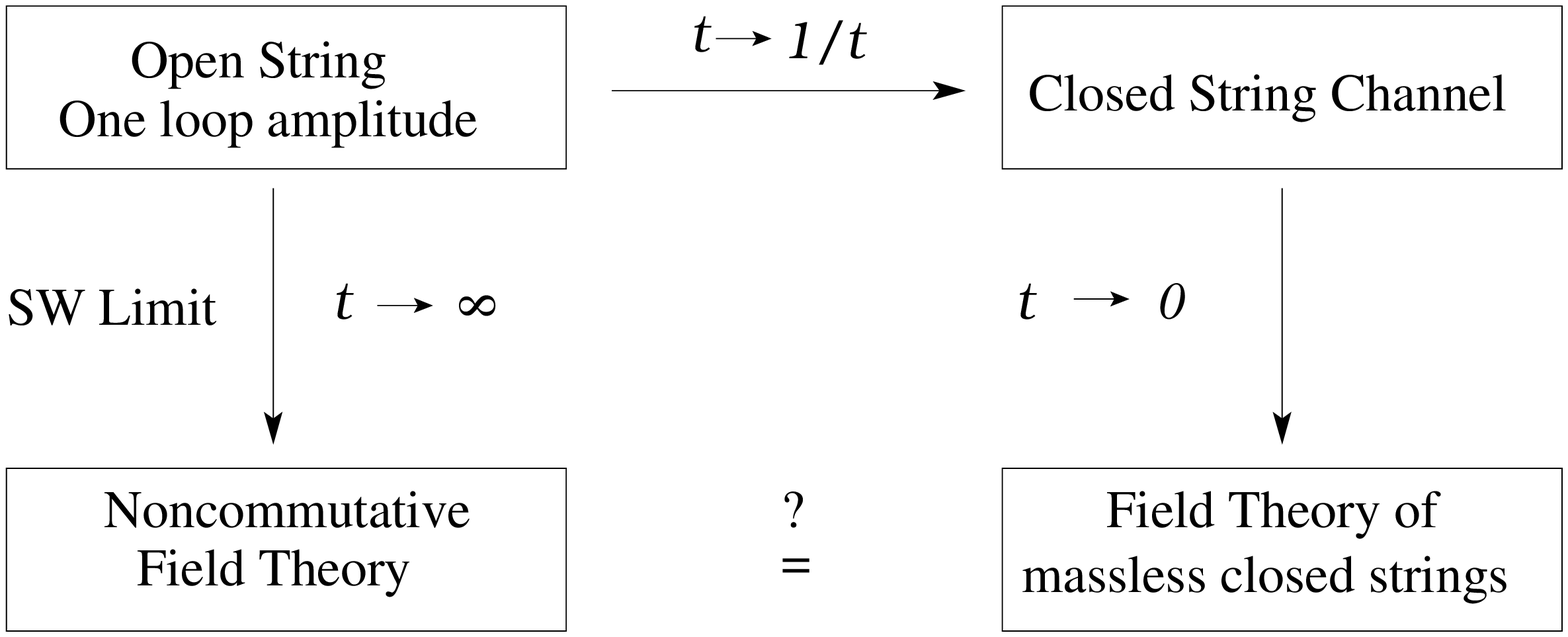, width= 10 cm,angle=0}
\vspace{ .1 in }
\caption[lim]{Noncommutative field theory and closed string channel
limits}
\end{center}
\end{figure}

\noindent
Figure 4 sums up the various limits involved 
in the problem addressed in this paper.
Noncommutative field theory arises in the Seiberg Witten limit. In the
open string one loop amplitude, the $t
\rightarrow \infty$ region of the moduli space of the cylinder
corresponds to the IR regime with only contributions to the
amplitude coming from the massless open string modes propagating in the
loop. As a result we get a one loop two point function in noncommutative 
field
theory. The other limit  $t \rightarrow 0$, corresponds to the UV region
of the open string loop. The one loop open string in this limit 
factorises
in the closed string channel. The contributions in this region come 
from the
massless tree level exchanges of the closed string modes. We have 
discussed in section 3. that the divergences arising from the two ends 
are related to each other (upto some overall normalisation). This 
relation could not be made exact in the setup considered here due to the 
presence of tachyons, which act as additional sources for divergences.
We have shown that the tensor structure for the noncommutative 
field theory (\ref{photon2pt}) two point amplitude can be recovered by 
considering 
massless and tachyonic exchanges of closed strings in the 
presence of background constant $B$-field. 
For the coefficient to match with the gauge theory result, in the 
bosonic string case, that we have studied,
the full tower of the closed string states are required. We expect that an 
exact correspondence 
between the UV behavior of the noncommutative gauge theory and massless 
closed string exchanges may be made in some compactified 
superstring theory, where the gauge theory is 
four dimensional 
and the closed strings move in exactly two extra transverse directions 
\cite{progress}.
This would cure the problem of tachyons as well as lead to the desired 
forms of propagators in the closed string channel. Keeping this 
in mind we have studied 
massless closed string exchanges in the background $B$-field. Apart from 
this it is an interesting 
problem by itself. The full two point 
amplitude in the presence of background $B$-field must be of the form 
(\ref{clchannel}). We have reconstructed this from the sum of massless 
(graviton, dilaton and $b$-field) exchanges with the vertices computed 
from the DBI action, by considering expansions of the amplitude in three 
different cases. This exercise has helped in isolating the contributions 
from each of the massless closed string modes separately. On the broader 
side 
this is one of the steps in the chain of limits in Figure 4.
We can view these results in the light of open/closed string duality, as 
quantities in the boundary noncommutative gauge theory are being 
recovered from the bulk theory of closed strings. In the usual case 
there are infinities on both sides, manifested as UV in 
one and IR in the other. The $B$-field in the SW limit acts as a 
background where at 
least a subsector (nonplanar) of the gauge theory is 
regularised. This 
is the sector that has been the subject in this paper. It would be 
interesting to find a limit where only the nonplanar sector of the 
noncommutative gauge theory survives.

}

\section{Open strings in background $B$-field}

In this section we give a short review of open string dynamics in the 
presence of constant background $B$-field leading to noncommutative 
field theory on the world volume of a $D$-brane \cite{Seiberg}. In the 
presence of a constant background $B$-field, the world sheet action is given by,

\beqa
S=\f{1}{4\pi\al} \int_{\Sigma} [g_{MN}\pa_{a}X^{M}\pa^{a}X^{N} -2\pi 
i\al 
B_{MN}\epsilon^{ab}\pa_{a}X^{M}\pa_{b}X^{N}]
\eeqa

\noindent
Consider a $D_p$ brane extending in the directions $1$ to $p$, such 
that, $B_{MN} \neq 0$ only for $M,N \leq p+1$ and $g_{MN}=0$ for $M \leq 
p+1, N > p$. The equation of motion gives the following boundary 
condition,

\beqa
g_{MN}\pa_{n}X^{N}+2\pi i \al B_{MN}\pa_{t} X^{N}|_{\pa\Sigma}=0
\eeqa

The world sheet propagator on the boundary of a disc satisfying this 
boundary condition is given by,

\beqa
{\cal G}(y,y^{'})=-\al 
G^{MN}\ln(y-y^{'})^2+\f{i}{2}\th^{MN}\ep(y-y^{'})
\eeqa

\noindent
where, $\ep(\Delta y)$ is $1$ for $\Delta y>0$ and $-1$ for 
$\Delta y<0$. $G_{MN}$, $\th_{MN}$ are given by,

\beqal{gth}
G^{MN}&=&\left(\f{1}{g+2\pi\al B}g\f{1}{g-2\pi\al B}\right)^{MN}\non
G_{MN}&=&g_{MN}-(2\pi\al)^2(Bg^{-1}B)_{MN}\non
\th^{MN}&=&-(2\pi\al)^2\left(\f{1}{g+2\pi\al B}B\f{1}{g-2\pi\al 
B}\right)^{MN}
\eeqa

The relations above define the open string metric $G$ in terms of the 
closed string metric $g$ and $B$. This difference in the two metrics as 
seen 
by the open strings on the brane and the closed strings in the bulk 
plays an important role in the discussions in the following sections.
We next turn to to the low energy limit, $\al \rightarrow 0$. A 
nontrivial low energy theory results from the following scaling.

\beqal{swl}
\al \sim \epsilon^{1/2} \rightarrow 0 
\mbox{\hspace{0.1in};\hspace{0.1in}} 
g_{ij} \sim \epsilon \rightarrow 0
\eeqa

where, $i,j$ are the directions along the brane. This is the 
Seiberg Witten (SW) limit which gives rise to noncommutative field 
theory on the brane. The relations in eqn(\ref{gth}), to the leading 
orders, in this 
limit reduce to,

\beqa
G^{ij}&=&-\f{1}{(2\pi\al)^2}(\th g \th)^{ij} 
\mbox{\hspace{0.1in};\hspace{0.1in}}
G_{ij}=-(2\pi\al)^2(Bg^{-1}B)_{ij}\non 
\th^{ij}&=&\left(\f{1}{B}\right)^{ij}
\eeqa    

for directions along the $D_p$ brane. $G_{MN}=g_{MN}$ and $\th=0$ 
otherwise. It was shown that the tree-level action for the low energy 
effective field theory on the brane has the following form,

\beqa
S_{YM}=-\f{1}{g_{YM}^2}\int \sqrt{G} 
G^{kk^{'}}G^{ll^{'}}\Tr(\hat{F}_{kl}*\hat{F}_{k^{'}l^{'}})
\eeqa

\noindent
where the $*$-product is defined by, 

\beqa
f*g(x)=e^{\f{i}{2}\th^{ij}\pa_{i}^y\pa_{j}^z}f(y)g(z)\mid_{y=z=x}
\eeqa

\noindent
and $\hat{F}_{kl}$ is the noncommutative field strength, which is 
related to the ordinary field strength, $F_{kl}$ by,

\beqal{redeff}
\hat{F}_{kl}=F_{kl}+\th^{ij}(F_{ki}F_{lj}-A_i\pa_{j}F_{kl})+{\cal 
O}(F^3)
\eeqa

\noindent
and,

\beqa
\hat{F}_{kl}=\pa_k\hat{A}_l-\pa_l\hat{A}_k-i\hat{A}_k*\hat{A}_l+
i\hat{A}_l*\hat{A}_k
\eeqa

Noncommutative field theories as defined here have been studied 
extensively for the last few years. 
One of the most important features of these theories is the coupling of 
the UV and the IR regimes, manifested by the nonplanar sector 
of these theories, contradicting our usual notions of Wilsonian 
RG \cite{minwalla}. This mixing of the UV and IR sectors also occurs in 
scalar theories, where the noncommutative version is formulated by 
replacing all products of fields by $*$-products. 
We write down here a simple two point nonplanar one 
loop amplitude 
for noncommutative $\phi^4$ theory in four dimensions, in the continuum 
limit,
  
\beqal{scalar2pt}
\Gamma^2(p) \sim \f{1}{\tilde{p}^2}-m^2\ln(\tilde{p}^2m^2)
\eeqa

\noindent
where, $\tilde{p}=(\th p)$. 
The amplitude is finite in the UV but is IR divergent though we had a 
massive theory to start 
with. Note that $\tilde{p}^2$ plays the role of $1/\Lambda^2$, where 
$\Lambda$ is the UV cutoff. It 
was suggested \cite{minwalla} that these IR divergent terms could 
arise 
by integrating out massless modes at high energies. This is quite 
like the open string one loop divergence which is reinterpreted as IR 
divergence coming from massless closed string exchange. It was noted 
that the first and second terms of eqn(\ref{scalar2pt}) can be recovered 
through massless 
tree-level exchanges if these modes are allowed to propagate in $0$ and 
$2$ extra dimensions transverse to the brane respectively 
\cite{raamsdonk}. A similar 
structure arises for the nonplanar two point function for the gauge 
boson in noncommutative gauge theories,

\beqal{photon2pt}
\Pi^{ij}(p) \sim 
N_1[G^{ij}G^{kl}-G^{ik}G^{jl}]p_kp_l\ln(p^2\tilde{p}^2) 
+N_2 \f{\tilde{p}^{i}\tilde{p}^{j}}{\tilde{p}^4}
\eeqa

\noindent
$N_1$ and $N_2$ depends on the matter content of the theory. For some 
early works on noncommutative gauge theories see \cite{martin}.
The effective action with the two point function (\ref{photon2pt}) is 
not gauge invariant.
To write down a gauge invariant effective action one needs to introduce 
open Wilson lines \cite{kawai} 

\beqa
W_C(p)=\int d^4x P*\exp\left(ig\int_C d\s \pa_{\s}y^{i} 
A_{i}(x+y(\s)\right)*e^{ipx}
\eeqa

\noindent
The curve $C$ is parametrized by $y^{i}(\s)$, where $0\leq \s \leq 1$ 
such that, $y^{i}(1)-y^{i}(0)=\tilde{p}^{i}$. 
Correlators of Wilson lines in noncommutative gauge theories have been 
studied by various authors \cite{wilsonline}. The terms in 
(\ref{photon2pt}) are the leading terms in the expansion of 
the two point function for the open Wilson line.
A crucial point to be noted is that for supersymmetric theories, $N_2$, 
the coefficient of the second term, which is allowed by the 
noncommutative gauge invariance vanishes \cite{matusis}. 
Also see \cite{khoze} for an elaborate discussion. An 
observation on the arising of tachyon in the closed string theory 
in the bulk and the non vanishing of $N_2$ was made in \cite{armoni}. 
Attempts have 
been made, along the lines as discussed above, to recover the nonplanar
IR divergent terms from tree-level closed string exchanges. We analyse 
this issue in the next section.

\section{Open string one loop amplitude}

In this section we compute the open string one loop amplitude with 
insertion of two gauge field vertices. We will compute the two point 
amplitude in the closed string channel keeping only the contributions 
from the tachyon and the massless modes. One loop amplitudes for open 
strings with two vertex insertions in the presence of a constant 
background $B$-field have been computed by various 
authors, and field theory amplitudes were obtained in the 
$\al\rightarrow 0$ limit \cite{oneloop}. 

\noindent
Firstly, the one loop partition function is written as 
\cite{polchinski,callan} 

\beqa
Z(t)=\det(g+2\pi\al B)Z_0(t)
\eeqa

with,
\beqa
Z_0(t)= \Tr[\exp(-2\pi t L_0)]
\eeqa

where, $t$ is the modulus of the cylinder and $L_0$ is given by,

\beqa
L_0=\al G_{ij}k^{i}k^{j}+\f{(X^{\alpha})^2}{4\pi^2\al}+\f{1}{2} 
\sum _{q\neq 0}G_{MN}a_{q}^{M}a_{-q}^{N}
\eeqa

For an open string ending on a $D_p$ brane ($X^{\alpha}=0$), this gives,

\beqal{zt}
Z(t)=\det(g+2\pi\al B){\cal V}_{p+1}(8\pi^2\al 
t)^{-\f{p+1}{2}}\eta(it)^{-(D-2)}
\eeqa

\noindent
${\cal V}_{p+1}$ is the volume of the $D_p$ brane.
We are interested in the two point one loop amplitude. 
Specifically we write down here 
the nonplanar amplitude for reasons mentioned earlier. 
The two point one loop amplitude has the form,

\beqa
A(p_1,p_2)=\int_{0}^{\infty}\f{dt}{2t}Z(t)\int_{0}^{2\pi 
t}dy\int_{0}^{2\pi t}dy^{'} 
<V(p_1,x,y)V(p_2,x^{'},y^{'})>
\eeqa

\noindent
where $Z(t)$ is as defined in eqn(\ref{zt}). The required vertex 
operator is given 
by, 

\beqa
V(p,y)=-i\f{g_o}{(2\al)^{1/2}}\ep_{j}\pa_y 
X^{j}e^{ip.X}(y)
\eeqa

\noindent
The noncommutative field theory results are recovered from region of the 
modulus where $t\rightarrow \infty$ in the SW limit. As mentioned, the 
nonplanar diagrams in the noncommutative field theory gives rise to 
terms which manifest coupling of the UV to the IR sector of the 
field theory. 

\noindent
The $t \rightarrow 0$ limit, picks out the contributions only 
from the tree-level massless closed string exchange. This is the UV 
limit of the open string. The amplitude is usually divergent. However, 
in the usual case, these divergences are reinterpreted as IR divergences 
due to the 
massless closed string modes. What is the role played by the $B$-field? 
In the presence of the background $B$-field, the integral over the 
modulus is regulated. In the closed string side, this would mean that 
the propagator for the massless modes are modified so as to remove the 
IR divergences. We would now like to investigate this end of the 
modulus. 

Before going into the actual form let us see heuristically what 
we can expect to compare on both ends of the modulus.
First consider the one loop amplitude,

\beqal{op}
{\cal Z} \sim \int \f{dt}{t} (\al t)^{-\f{p+1}{2}} \eta(it)^{-(D-2)} 
\exp(-C/\al t)
\eeqa

where $C$ is some constant which in our case is dependent on the 
$B$-field. In the $t \rightarrow \infty$ limit,

\beqal{opc}
{\cal Z}_{op} \sim \int \f{dt}{t} (\al t)^{-\f{p+1}{2}} 
\left[e^{2\pi t}+(D-2)+O(e^{-2\pi t})\right]\exp(-C/\al t)
\eeqa

If we throw out the tachyon, and restrict ourselves only to the $O(1)$ 
term in the expansion of the $\eta$-function, $\al$ and $t$ occur in 
pairs. This means that in the $\al \rightarrow 0 $ limit the finite 
contributions to the field theory come from the region where $t$ is 
large. We can break the integral over $t$ into two parts, 
$1/\Lambda^2\al<t<\infty$ and $0<t<1/\Lambda^2\al$, where $\Lambda$ 
translates into the UV cutoff for the field theory on the brane. The 
second interval is the source of divergences in the field 
theory that is regulated by $C$. This is the region of 
the modulus dominated by massless exchanges in the closed string 
channel. For the closed string channel, we have

\beqal{clc}
{\cal Z}_{cl} \sim \int ds (\al)^{-\f{p+1}{2}}s^{-l/2}
\left[e^{2\pi s}+(D-2)+O(e^{-2\pi s})\right]\exp(-Cs/\al )
\eeqa

where $l=D-(p+1)$, is the number of dimensions transverse to the $D_p$ 
brane. The would be divergences as $C \rightarrow 0$ manifest themselves 
as $1/C$ or $\ln(C)$, depending on $l$ \cite{raamsdonk}. The full open 
string channel 
result will always require all the closed string modes for its dual 
description. As far as the divergent (UV/IR mixing) terms are 
concerned, we can hope to realise them through some field theory of the 
massless closed string modes. However, the exact correspondence between the 
divergences in both the channels, is destroyed by the presence of the 
tachyons. Also note that, at the $t \rightarrow 0$ end of 
the open 
string one loop amplitude, the divergence 
is contributed by the full
tower of open string modes. However, we want that the one loop open
string amplitude restricted to only the
massless exchanges to be rewritten as massless closed string exchanges.
For this to happen the integrand as a function of $t$ in one loop
amplitude should have the same asymptotic form as $t\rightarrow 0$ and
$t \rightarrow \infty$ so that eqn(\ref{clc}) is exactly the same as 
that of eqn(\ref{opc}) 
integrated between $[0,1/\Lambda^2\al]$. 
There are examples of supersymmetric configurations where the one
loop open string amplitude restricted to the massless sector can be
rewritten exactly as tree-level massless closed string exchanges. It was 
shown that in these situations the potential between two branes
with separation $r$ is the same at both the $r\rightarrow 0$, and
$r\rightarrow \infty$ corresponding to
$t\rightarrow \infty$ and $t\rightarrow 0$ ends
respectively \cite{douglas2}. Consequences of this fact in relation 
to the gauge/gravity correspondence have 
been explored in \cite{DiVecchia}. We can
expect that in these cases the IR singularities of the noncommutative
gauge theory match with those computed from the closed string massless
exchanges. In the bosonic case, this is true
for $p=13$, if we remove the tachyons. However, we are concerned with 
reproducing UV/IR effects of 
four dimensional gauge theory for which we need to set $l=2$.
The broader purpose of the exercise that follows is to outline a 
construction that can be set up for supersymmetric cases.  

We now return to the original computation of the amplitude in the closed
string channel. The nonplanar world sheet propagator obtained by 
restricting to the 
positions at the two boundaries is \cite{callan,oneloop},  

\beqa
{\cal G}^{ij}(y,y^{'})=-\al 
G^{ij}\ln\left|e^{-\f{\pi}{4t}}
\f{\vth_4\left(\f{\Delta y}{2\pi t},\f{i}{t}\right)}
{t^{-1}\eta(i/t)^3}\right|^2
-i\f{\th^{ij}\Delta y}{2\pi t}
-\al g^{ij}\f{\pi}{2t}
\eeqa

\noindent
where, $\Delta y=y-y^{'}$.
In the limit $t\rightarrow 0$ the propagator has the following 
structure,
  
\beqa
{\cal G}^{ij}=-4\al G^{ij}
\left[cos(\Delta y/t)e^{-\f{\pi}{t}}-e^{-\f{2\pi}{t}}\right]
-i\f{\th^{ij}\Delta y}{2\pi t}
-\al g^{ij}\f{\pi}{2t}
\eeqa

\noindent
Inserting this into the correlator for two gauge bosons and keeping 
only terms that would contribute to the tachyonic and massless closed 
string exchanges, we get, 

\beqa
<...>=\left[p_{k}p_{l}\f{(8\pi\al)^2}{(2\pi t)^2}
(G^{ij}G^{kl}-G^{ik}G^{jl})sin^2(\Delta 
y/t)e^{-\f{2\pi}{t}}+\f{\tilde{p}^{i}\tilde{p}^{j}}{(2\pi t)^2}
\right]e^{p_{i}{\cal G}^{ij}p_{j}}
\eeqa

expanding $\eta(it)$ in this limit,

\beqa
\eta(it)^{-(D-2)}=t^{\f{D-2}{2}}\eta(i/t)^{-(D-2)} &\sim& 
t^{\f{D-2}{2}}[e^{\f{2\pi}{t}}+(D-2) +O(e^{-\f{2\pi}{t}})]
\eeqa

The two point amplitude with only the tachyonic and the massless closed 
string exchange can now be written down,

\beqa
A_2(p,-p)=-i\det(g+2\pi\al B){\cal V}_{p+1}(\f{g_o^2}{2\al})
(8\pi^2\al )^{-\f{p+1}{2}}\ep_{i}\ep_{j}I(p)
\eeqa

\noindent
with $I(p)=I_{T}(p)+I_{\chi}(p)$ and,

\def\note{Note that the $s$ integral has to be cutoff at the lower end 
at some value 
$\Lambda^2\al$. This corresponds to the UV transverse momentum cutoff 
for the closed strings, that allows us to extract the contribution from 
the IR region.
\beqal{cutoff}
I(p,\Lambda)\sim 
\int_{\Lambda^2\al}^{\infty}\f{ds}{s}e^{-p^2\al s}
\sim \int_{0}^{\infty}d^2\kpe 
\f{e^{-(\kpe^2+p^2)\Lambda^2\als}}{(\kpe^2+p^2)\al}
\eeqa
The integral over $\kpe$, eqn(\ref{cutoff}) receives contribution upto 
$\kpe\sim 
O(1/\Lambda\al)$.
The included region of the $\kpe$ integral is the required IR sector for 
the transverse closed string modes or the UV for the open string 
channel.
}

\beqal{T}
I_{T}(p)&=&\tilde{p}^{i}\tilde{p}^{j}\int 
ds s^{-\f{l}{2}}\exp\left\{-(\f{\al\pi}{2}p_{i}g^{ij}p_{j}
-2\pi)s\right\}\\
&=&4\pi(2\pi^2\al)^{\f{l}{2}-1}\tilde{p}^{i}\tilde{p}^{j}\int 
\f{d^l\kpe}{(2\pi)^l}\f{1}{\kpe^2+p_{i}g^{ij}p_{j}-4/\al}
\nonumber
\eeqa

We have written the integral over $t$ in terms of $s=1/t$ in (\ref{T}) 
and further in the last expression we have replaced the integral over 
$s$ with that of $\kpe$.
The dimension of the $\kpe$ integral is the number of directions 
transverse to the brane and is thus the momentum of the closed string along 
these directions.
\note{}With this observation, for the tachyon with $l=2$, we get

\beqal{tac2pt}
I_{T}(p,\Lambda)&=&4\pi^2(2\pi^2\al)^{\f{l}{2}-1}\tilde{p}^{i}\tilde{p}^{j}
\ln\left(\f{p_{i}g^{ij}p_{j}-4/\al+\f{1}{(\Lambda\al)^2}}
{p_{i}g^{ij}p_{j}-4/\al}\right)
\eeqa

\noindent
For the noncommutative limit (\ref{swl}), we can expand the answer 
(\ref{tac2pt}) in powers of 
$1/(\al pg^{-1}p)$,

\beqal{texp}
\ln\left(pg^{-1}p-4/\al\right)\sim 
\ln\left(pg^{-1}p\right)-\f{4}{\al pg^{-1}p}-
\f{1}{2}\left(\f{4}{\al pg^{-1}p}\right)^2 - \textellipsis
\eeqa

\noindent
The $(1/\al pg^{-1}p)^2$ term in the expansion 
(\ref{texp}) above  
corresponds to the IR singular term which appears in the noncommutative 
gauge theory. To compare with the second term of
(\ref{photon2pt}), we should set $G=\eta$, so that $g^{-1} \sim 
-\th^2/\als $.
Here we have got this from one of the 
terms in the expansion of the amplitude with tachyon 
exchange. However 
one can easily see that any massive spin zero closed string exchange 
would produce such a term. As far as the exact coefficient is 
concerned, the full tower of massive states would contribute. 
The absence of this term in the supersymmetric theories can only be due 
to exact cancellations between the bosonic and fermionic sector 
contributions \cite{armoni}.

As for the tachyon, similarly we now write down the contribution from 
the massless exchanges,

\beqal{clchannel}
I_{\chi}(p,\Lambda)&=&4\pi(2\pi^2\al)^{\f{l}{2}-1}\left[(D-2)
\tilde{p}^{i}\tilde{p}^{j}+8(2\pi\al)^2 p_{k}p_{l}
(G^{ij}G^{kl}-G^{ik}G^{jl})\right]\times\non
&\times&
\int
\f{d^l\kpe}{(2\pi)^l}\f{1}{\kpe^2+p_{i}g^{ij}p_{j}}
\eeqa

\noindent
One can observe that the terms occurring with $\als (\sim \ep)$ as the 
coefficient, relative to the other terms in
(\ref{texp}) and (\ref{clchannel}), appear 
in the gauge theory result in eqn(\ref{photon2pt}).
In the closed string channel we have got this for the number of 
transverse dimensions, $l=2$. This means that $p+1=D-2=24$ is the 
dimension of the gauge theory on the string side. However the result of 
eqn(\ref{photon2pt}) is valid for the NC gauge theory defined in 
4-dimensions.
To understand why it is these terms that occur in the four dimensional 
gauge theory, we must have a string setting where $l=2$ and $p=3$.
However, at this point, as discussed earlier, it is only necessary that 
$l=2$ so 
that the lowest lying closed string exchanges reproduce the
correct form of the IR singularities as that of the gauge theory in 
eqn(\ref{photon2pt}).

We mention again that the exact correspondence between the UV behavior 
of the noncommutative gauge theory and closed string exchanges 
would require the full tower of closed string states. The contribution 
from the massive closed string states are likely to be supressed only in 
some supersymmetric configurations \cite{douglas2}\cite{DiVecchia}. 
Keeping in mind these situations we compute the exchanges due to 
massless closed strings in 
the presence of background $B$-field in the following section.

%\noindent
%The full normalisation coefficient for the closed string exchanges is,
%\beqal{norm}
%{\cal N}=-i\det(g+2\pi\al B){\cal V}_{p+1}\left(\f{g_o^2}{2\al}\right)
%\f{\pi(4\pi^2\al)^{(D-2)/2-(p+1)}}{2^{(D-2)/2-2}}
%\eeqa

\section{Closed string exchange}

In this section we reconstruct the two point function of two gauge 
fields eqn(\ref{clchannel})
with massless closed string exchanges. The aim here is to write the 
amplitude as sum of massless closed string exchanges in the presence of
constant background $B$-field.
To proceed, by considering the effective field theory of massless
closed strings, we construct the propagators for these modes 
(graviton, dilaton and $B$-field) with a constant background $B$-field. 
As a next step we compute the couplings of 
the gauge field on the brane with the massless closed strings from the 
DBI action. Finally we combine these results to construct the two point 
function. We will consider three separate cases when computing the two 
point amplitude in this section. (I) In this case the background $B$ 
field is assumed to be small and the closed string metric, $g=\eta$.
(II)The Seiberg Witten limit when $g=\ep\eta$. (III)The case when the 
open 
string metric on the brane, $G=\eta$. The amplitude eqn(\ref{clchannel}) 
in the closed 
string channel is the closed form result of the massless exchanges. In 
each of the above cases, we will compare this amplitude to respective
orders with the ones we compute here in this section.
Let us first begin by considering the field theory of the massless modes 
of the closed string string propagating in the bulk. The spacetime action for 
closed string fields is written as,

\beqa
S=\f{1}{2\ka^2}\int d^DX
\sqrt{-g}[R-\f{1}{12}
e^{-\f{8\phi}{D-2}}H_{LMN}H^{LMN}
-\f{4}{D-2}g^{MN}\pa_{M}\phi\pa_{N}\phi]
\eeqa

Where $D$ is the number of dimensions in which the closed string
propagates. The indices are raised and lowered by $g$. We will now 
construct the tree-level propagators that will 
be necessary in the next section to compute two point amplitudes. For 
each of the cases as defined above, the propagator will take a 
different form. Let us first consider the dilaton. For $g=\eta$ the 
propagator is the usual one,

\beqal{dp1}
<\phi\phi>&=&-\f{(D-2)i\ka^2}{4}\f{1}{\kpe^2+\kpa^2}
\eeqa

The next limit for the metric is $g=\ep\eta$ along the world volume 
directions of the brane. In this limit, the dilaton 
part of the action can be written as,

\beqa
S_{\phi}=-\f{4}{\kappa^2(D-2)}\int d^DX
\f{1}{2}[\pa_{\alpha}\phi\pa^{\alpha}\phi+\ep^{-1}
\pa_{i}\phi\pa^{i}\phi]
\eeqa

This gives the propagator,

\beqal{dp2}
<\phi\phi>&=&-\f{(D-2)i\ka^2}{4}\f{1}{\kpe^2+\ep^{-1}\kpa^2}
\eeqa

Finally, when the open string metric is set to, $G=\eta$,
the lowest order solution for $g$ along the brane directions is,

\beqa
g=-(2\pi\al)^2B^2 +{\cal O}(\alpha^{'4})
\eeqa

which gives,

\beqal{dp3}
<\phi\phi>&=&-\f{(D-2)i\ka^2}{4}\f{1}{\kpe^2+\tilde{\kpa}^2/(2\pi\al)^2}
\eeqa

where,
\beqa
\tilde{\kpa}^2=-\kpai\left(\f{1}{B^2}\right)^{ij}\kpaj
\eeqa

Let us now turn to the free part for the antisymmetric tensor field,

\beqa
S_b=-\f{1}{24\kappa^2}\int d^DX H_{LMN}H^{LMN}
\eeqa

where,
\beqa
H_{LMN}=\pa_{L}b_{MN}+\pa_{M}b_{NL}
+\pa_{N}b_{LM}
\eeqa

Using the following gauge fixing condition,

\beqa
g^{MN}\pa_{M}b_{NL}=0
\eeqa

The action reduces to,

\beqa
S_b=-\f{(2\pi\al)^2}{8\kappa^2}\int d^DX 
\left[g^{\alpha\beta}\pa_{\alpha}
b_{IJ}\pa_{\beta}b_{KL}+g^{ij}\pa_i b_{IJ}\pa_j 
b_{KL}\right]g^{IK}g^{JL}
\eeqa

The factor of $(2\pi\al)^2$ in the $b$-field action has been 
included because
the sigma model is defined with $(2\pi\al)B$ coupling.
The propagator then is,

\beqal{bp}
<b_{IJ}b_{I^{'}J{'}}>=-\f{2i\ka^2}{(2\pi\al)^2}\f{g_{I[J{'}}g_{I^{'}]J}}
{\kpe^2+g^{ij}\kpai\kpaj}
\eeqa

Finally, the gravitational part of the action. As will turn out in the 
next section that we will only have to consider graviton exchanges for 
the case $g=\eta$. The propagator for the graviton here is the usual 
propagator from the action,

\beqa
S_h=\f{1}{2\ka^2}\int d^DX
\sqrt{-g}R
\eeqa

By considering fluctuations about $\eta$, and in the gauge 
(\ref{hgauge}),

\beqa
g_{MN}=\eta_{MN}+h_{MN} 
\eeqa

\beqal{hgauge}
g^{MN}\Gamma^{L}_{MN}=0
\eeqa

the graviton propagator is,

\beqal{gp}
<h_{IJ}h_{I^{'}J{'}}>=-2i\ka^2\f{[\eta_{I\{J{'}}\eta_{I^{'}\}J}-2/(D-2)
\eta_{IJ}\eta_{I^{'}J{'}}]}{\kpe^2+\kpa^2}
\eeqa

After writing down the required propagators, we now turn to the 
computation of the vertices. As mentioned in the beginning of this 
section, we will consider each of the three cases separately. To begin,
we first write down the DBI action for a $D_p$ brane,

\beqa
S_p=-T_p\int d^{p+1}\xi e^{-\Phi}\sqrt{g^{'}+2\pi\al(B+b)}
\eeqa

Where, $g^{'}$ is the closed string metric in the string frame, $B$ is 
the constant two form
background field and $b$ is the fluctuation of the two form field. The
$b$-field on the brane is interpreted as the two form field strength 
for
the $U(1)$ gauge field and in the bulk it is the usual two form 
potential. Going to the Einstein frame by defining,

\beqa
g =g^{'}e^{2\omega}\mbox{;\hspace{0.2in}}
\omega=\f{2(\phi_0-\Phi)}{D-2}\mbox{;\hspace{0.2in}}
\Phi=\phi+\phi_0 \mbox{;\hspace{0.2in}}
\omega=\f{-2\phi}{D-2}
\eeqa

the action can be rewritten as,

\beqa
S_p=&-&\tau_p\int d^{p+1}\xi {\cal L}(\phi,,h,b)\non
&-&\tau_p\int d^{p+1}\xi e^{-\phi(1-\f{2(p+1)}{D-2})}
\sqrt{g+2\pi\al(B+b)e^{-\f{4\phi}{D-2}}}
\eeqa

where, $\tau_p=T_pe^{-\phi_0}$ and $\phi$ is the propagating dilaton
field. We will now consider each of the three cases separately and 
compute the two point function upto the respective orders.

\subsection{Expansion for small $B$}

In this part we compute the couplings of the gauge field on the brane to 
the massless closed strings in the bulk. We will assume the background 
constant $B$-field to be small and compute the lowest order 
contribution to the two point function considered as an expansion in 
$B$. The first thing to note is that, since $B$ is antisymmetric,
there cannot be a non vanishing amplitude with a single $B$ in one vertex
only. We need at least two powers of $B$. 
One on each vertex or both on one. The graviton and the dilaton
need one on each vertex. The $b$-field  can couple to the gauge field
without a $B$. So for the $b$-field we need to consider couplings upto
${\cal O}(B^2)$. The closed string tree-level diagrams contributing to 
the three massless modes are shown in Figure 1.

\beqa
{\cal L}=\sqrt{e^{-P\phi}\left[g+(2\pi\al)e^{-Q\phi}(B+b)\right]}
\eeqa

\noindent
where
\beqa
P=\f{2}{p+1}-\f{4}{(D-2)} \mbox{\hspace{0.2in}} Q=\f{4}{(D-2)}
\eeqa

We now expand of ${\cal L}$ for small $B$, with $g=\eta+h$,

\beqal{exp}
{\cal L}=\sqrt{e^{-P\phi}[g+(2\pi\al)e^{-Q\phi}b]}
\left[1+\f{(2\pi\al)e^{-Q\phi}}{g+(2\pi\al)e^{-Q\phi}b}B\right]^{1/2}
\eeqa

To the linear order in $B$,
\beqa
{\cal L}
&=&\sqrt{e^{-P\phi}[g+(2\pi\al)e^{-Q\phi}b]}\left[1+\f{(2\pi\al)}{2}
e^{-Q\phi}\Tr\f{1}{g+(2\pi\al)e^{-Q\phi}b}B\right]\non
&=&\sqrt{e^{-P\phi}[g+(2\pi\al)e^{-Q\phi}b]}
\left[1-\f{1}{2}(2\pi\al)^2e^{-2Q\phi}\Tr\left\{g^{-2}bB\right\}
\right]
\eeqa

In the last line the trace was expanded in powers of $\al$. The first 
term is zero because it is trace over an antisymmetric matrix.
Let us define the term under the square-root in the last line as $Y$ and 
the second term as $X$,

\beqa
Y&=&\sqrt{e^{-P\phi}[\eta+h+(2\pi\al)e^{-Q\phi}b]}\\
X&=&\left[1-\f{1}{2}(2\pi\al)^2e^{-2Q\phi}\Tr\left\{(\eta+h)^{-2}bB\right\}
\right]
\eeqa

To get the vertices, we need to find,

\beqa
\f{\de^2{\cal L}}{\de b\de \chi}&=&\f{\de^2 (XY)}{\de b\de \chi}\\
&=&\left[X\f{\de^2Y}{\de b\de \chi}+\f{\de Y}{\de \chi}\f{\de X}{\de 
b} +\f{\de X}{\de \chi}\f{\de Y}{\de b}+ Y \f{\de^2X}{\de b\de \chi}
\right]
\eeqa
where, $\chi \equiv \phi,b,h$

\noindent
Now, listing the required derivatives at $\phi,b,h=0$,

\beqa
\f{\de X}{\de h_{ij}}=0 \mbox{\hspace{.15in};\hspace{.15in}}
\f{\de X}{\de b_{kl}}=-\f{1}{2}(2\pi\al)^2B^{lk}
\mbox{\hspace{.15in};\hspace{.15in}}
\f{\de X}{\de \phi}=0
\eeqa
\beqa
\f{\de^2 X}{\de b_{kl} \de h_{ij}}=(2\pi\al)^2\eta^{jk}B^{li}
\mbox{\hspace{.05in};\hspace{.05in}}
\f{\de^2 X}{\de b_{kl} \de \phi}=(2\pi\al)^2QB^{lk}
\mbox{\hspace{.05in};\hspace{.05in}}
\f{\de^2 X}{\de b_{kl} \de b_{ij}}=0
\eeqa
\beqa
\f{\de Y}{\de h_{ij}}=\f{1}{2}\eta^{ij}
\mbox{\hspace{.15in};\hspace{.15in}}
\f{\de Y}{\de b_{kl}}=0
\mbox{\hspace{.15in};\hspace{.15in}}
\f{\de Y}{\de \phi}=-\f{p+1}{2}P
\eeqa
\beqa
\f{\de^2 Y}{\de b_{kl} \de h_{ij}}=0
\mbox{\hspace{.05in};\hspace{.05in}}
\f{\de^2 Y}{\de b_{kl} \de \phi}=0
\mbox{\hspace{.05in};\hspace{.05in}}
\f{\de^2 Y}{\de b_{kl} \de b_{ij}}=(2\pi\al)\eta^{li}\eta^{jk}
\eeqa

Using these derivatives, the vertices for the graviton and dilaton are,

\beqa
V_h^{ij}&=&-\tau_p(2\pi\al)^2\left[-\f{1}{4}B^{lk}\eta^{ij}
+\eta^{jk}B^{li}\right]\\
V_{\phi}&=&-\tau_p(2\pi\al)^2\left[\f{1}{4}(p+1)P+Q\right]B^{lk}
\eeqa

For the $b$-field we need to consider couplings upto 
${\cal O}(B^2)$, the next order term in $B$ in the expansion of 
eqn(\ref{exp}). Since we are not interested in the graviton and dilaton 
exchange at this order, so  putting them to zero,

\beqa
{\cal 
O}(B^2)&=&\sqrt{\eta+(2\pi\al)b}\left[-\f{1}{4}\Tr\left(\f{(2\pi\al)}
{\eta+(2\pi\al)b}B\right)^2+\f{1}{8}\left(\Tr\f{(2\pi\al)}
{\eta+(2\pi\al)b}B\right)^2\right]\non
&=&(2\pi\al)^2\left(1-\f{(2\pi\al)^2}{4}\Tr(b^2)\right)\times\\
&\times&\left[-\f{1}{4}\Tr\left(
B^2+(2\pi\al)^2(bBbB 
+2b^2B^2)\right)+\f{(2\pi\al)^2}{8}\Tr(bB)\Tr(bB)\right]\nonumber
\eeqa

This along with the ${\cal O}(1)$ term gives the following vertex
for the $b$-field.

\beqa
V_b^{ij}&=&\tau_p\f{(2\pi\al)^2}{2}\eta^{li}\eta^{jk}
\left(1-(2\pi\al)^2\f{1}{4}\Tr(B^2)\right)
\non
&-&\tau_p(2\pi\al)^4\left[\f{1}{4}B^{kl}B^{ij}-\f{1}{2}B^{li}B^{jk}
-(B^2)^{li}\eta^{jk}\right]
\eeqa

\begin{figure}[t]
\begin{center}
\epsfig{file=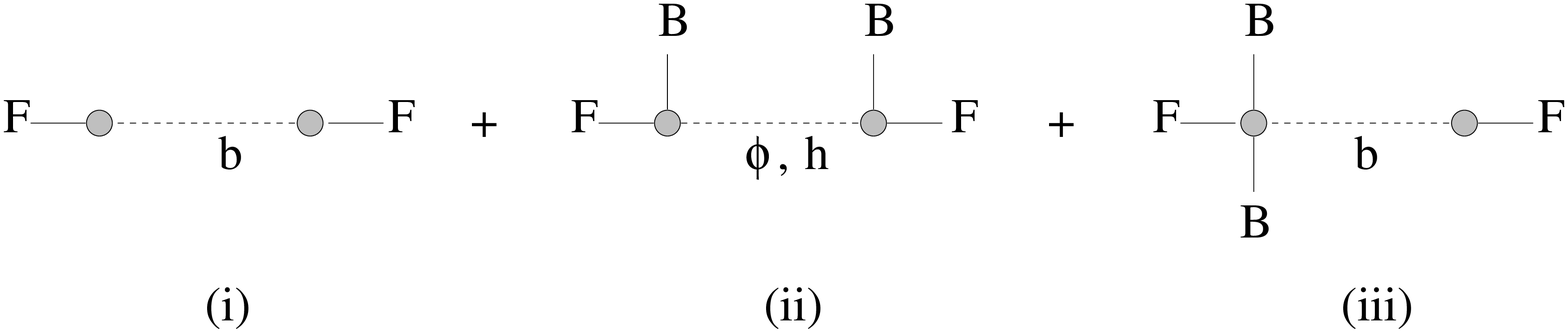, width= 12 cm,angle=0}
\vspace{ .1 in }
\begin{caption}
{Two point amplitude upto quadratic order in $B$. (i) and (iii) are due 
to 
only 
$b$-field exchange, (ii) is due to graviton and dilaton exchange.}
\end{caption}
\end{center}
\label{insertion}
\end{figure}

The propagators are the usual ones, rewriting them from eqns(\ref{dp1},
\ref{bp},\ref{gp}),

\beqa
<h_{ij}h_{i^{'}j^{'}}>&=&-2i\ka^2\f{[\eta_{ii^{'}}\eta_{jj^{'}}+
\eta_{ij^{'}}\eta_{i^{'}j}-2/(D-2)\eta_{ij}\eta_{i^{'}j^{'}}]}{\kpe^2+\kpa^2}\\
<\phi\phi>&=&-\f{(D-2)i\ka^2}{4}\f{1}{\kpe^2+\kpa^2}\\
<b_{ij}b_{i^{'}j^{'}}>&=&-\f{2i\ka^2}{(2\pi\al)^2}\f{[\eta_{ii^{'}}\eta_{jj^{'}}-\eta_{ji^{'}}
\eta_{ij^{'}}]}{\kpe^2+\kpa^2}
\eeqa

\noindent
With these, the contributions from the three modes to the two point 
function can be worked out. 
We are interested in the correction to the quadratic term in the 
effective action for the 
gauge field on the brane. This can be constructed with the vertices 
computed above and the propagators for the intermediate massless closed 
string states. This correction for the nonplanar diagram can be written 
as, 

\beqal{pos2pt}
A_2(bb)=\int d^{p+1}\xi\int d^{p+1}\xi^{'}b(\xi)b(\xi^{'})
V<\chi(\xi)\chi(\xi^{'})>V
\eeqa
\noindent
where,
\beqa
<\chi(\xi)\chi(\xi^{'})>=\int 
\f{d^Dk}{(2\pi)^D}<\chi(\kpe,\kpa)\chi(-\kpe,-\kpa)>e^{-i\kpa(\xi-\xi^{'})}
\eeqa

\noindent
We can rewrite eqn(\ref{pos2pt}) in momentum space coordinates as,
 
\beqal{eff}
A_2(bb)&=&{\cal V}_{p+1}\int \f{d^{p+1}p}{(2\pi)^{p+1}}b(p)b(-p)\int 
\f{d^l\kpe}{(2\pi)^l}V<\chi(\kpe,-p)\chi(-\kpe,p)>V\non
&=&{\cal V}_{p+1}\int \f{d^{p+1}p}{(2\pi)^{p+1}}b(p)b(-p)L_2(p,-p)
\eeqa

\noindent
In the planar two point function, both the vertices are on the same end 
of the
cylinder in the worldsheet computation. In the field theory this 
corresponds to putting both the
vertices at the same position on the $D$-brane. In other words,
in the expansion of the DBI action, we should be looking for $b^2\chi$
vertices on one end and a $\chi$ tadpole on the
other. In this case, from the above calculation, $\kpa=0$. So 
the closed string propagator is just $1/\kpe^2$, i.e. the 
propagator is not modified by the momentum of the gauge field on the 
brane. This is what we expect, as in the field theory on the brane, the 
loop integrals are not modified for the planar diagrams. Here we will 
only concentrate on the nonplanar sector.
\\

\noindent
As mentioned earlier, on the brane we will identify, 
\beqal{ident}
b_{kl}(p)\equiv \f{g_0}{\sqrt{2\al}}F_{kl}(p)
=\f{g_0}{\sqrt{2\al}}p_{[k} A_{l]}(p)
\eeqa

For the graviton we have,

\beqa
L_2(bhb)&=&\int \f{d^l\kpe}{(2\pi)^l}
V_h^{ij,kl}<h_{ij}h_{i^{'}j^{'}}>V_h^{i^{'}j^{'},k^{'}l^{'}}\\
&=&-i\ka^2\tau_p^2 (2\pi\al)^4\int 
\f{d^l \kpe}{(2\pi)^l}\f{2}{\kpe^2+p^2}\times\\ 
&\times& \left[-B^{2ll^{'}}\eta^{kk^{'}}+B^{lk^{'}}B^{l^{'}k}+
\left(
\f{p+1}{8}+\f{p-1}{D-2}-\f{(p+1)^2}{8(D-2)}-1\right)
B^{lk}B^{l^{'}k^{'}}\right]\nonumber
\eeqa

\noindent
For the dilaton,

\beqa
L_2(b\phi b)&=&\int 
\f{d^l\kpe}{(2\pi)^l}V_{\phi}^{kl}<\phi\phi>V_{\phi}^{k^{'}l^{'}}\\
&=&-i\ka^2\tau_p^2 (2\pi\al)^4\int 
\f{d^l \kpe}{(2\pi)^l}\f{1}{\kpe^2+p^2}\f{(D-2)}{4}
\left(\f{1}{2}-\f{p+1}{D-2}+\f{4}{D-2}
\right)^2 B^{lk}B^{l^{'}k^{'}}\nonumber
\eeqa

\noindent
Adding the contributions from the graviton and the dilaton,
\beqal{ph}
L_2(bhb+ b\phi b)&=&-i\ka^2\tau_p^2(2\pi\al)^4\int
\f{d^l\kpe}{(2\pi)^l}\f{1}{\kpe^2+p^2}\times\\&\times&\left[
-2B^{2ll^{'}}\eta^{kk^{'}}+2B^{lk^{'}}B^{l^{'}k}+B^{lk}B^{l^{'}k^{'}}
\left(\f{D-2}{16}-1\right)\right]\nonumber
\eeqa

\noindent
Similarly for the $b$-field we have,
\beqal{b}
L_2(bbb)&=&\int
\f{d^l 
\kpe}{(2\pi)^l}V_b^{ij,kl}<b_{ij}b_{i^{'}j^{'}}>V_b^{i^{'}j^{'},k^{'}l^{'}}\\
&=&-i\ka^2\tau_p^2\int
\f{d^l \kpe}{(2\pi)^l}\f{1}{\kpe^2+p^2}\times\non&\times&
[\f{(2\pi\al)^2}{4}\{1-\f{(2\pi\al)^2}{2}\Tr(B^2)\}(\eta^{ll^{'}}\eta^{kk^{'}}-
\eta^{lk^{'}}\eta^{kl^{'}})\non
&+&(2\pi\al)^4\{\f{1}{2}B^{lk}B^{l^{'}k^{'}}+((B^2)^{ll^{'}}\eta^{kk^{'}}-
(B^2)^{lk^{'}}\eta^{kl^{'}})\non
&-&\f{1}{2}(B^{lk^{'}}B^{l^{'}k}-B^{ll^{'}}B^{k^{'}k})
+(lk)\leftrightarrow(l^{'}k^{'})\}]\nonumber
\eeqa

For the full two point function, there are cancellations between the 
eqn(\ref{ph}) and eqn(\ref{b}). The final answer is, 

\beqal{final1}
L_2&=&-i\ka^2\tau_p^2\int\f{d^l \kpe}{(2\pi)^l}
\f{1}{\kpe^2+p^2}\times\\&\times&[(2\pi\al)^4 
B^{lk}B^{l^{'}k^{'}}\f{D-2}{32}
+\f{(2\pi\al)^2}{4}\{1-\f{(2\pi\al)^2}{2}\Tr(B^2)\}(\eta^{ll^{'}}\eta^{kk^{'}}-
\eta^{lk^{'}}\eta^{kl^{'}})\non
&+&\f{(2\pi\al)^4}{2}\{(B^2)^{ll^{'}}\eta^{kk^{'}}-
(B^2)^{lk^{'}}\eta^{kl^{'}}\}+(lk)\leftrightarrow(l^{'}k^{'})]\nonumber
\eeqa

The full two point effective action, can now be constructed by putting 
back $L_2$ in eqn(\ref{eff}) along with the identification 
eqn(\ref{ident}).
To compare this with the closed string channel result with only massless 
exchanges, eqn(\ref{clchannel}) we must note the expansions of the 
following quantities to appropriate powers of $B$.

\beqa
G^{ij}&\sim & \eta^{ij}+(2\pi\al)^2(B^2)^{ij}+{\cal O}(B^4)\\
\th^{ij}&\sim &-(2\pi\al)^2B^{ij}+{\cal O}(B^3)\\
\sqrt{\eta+(2\pi\al)B}&\sim &\left[1-\f{(2\pi\al)^2}{4}\Tr(B^2)
+{\cal O}(B^4)\right]
\eeqa

With these expansions, we can see that eqn(\ref{clchannel}) equals the 
sum of massless contributions, in eqn(\ref{final1}). 

%The
%normalisation coefficient for eqn(\ref{eff}) is then,

%\beqa
%{\cal N}=-\f{i}{4}{\cal 
%V}_{p+1}\left(\f{g_0^2}{2\al}\right)\ka^2\tau_p^2
%\eeqa

\subsection{Noncommutative case $(g=\ep\eta)$}

We now turn to the Seiberg Witten limit, (\ref{swl}) which gives rise to 
noncommutative field theory on the brane. Here again we will be 
interested in writing out the two point function eqn(\ref{clchannel}) in 
the closed string 
channel as a sum of the massless closed string modes. Due to the scaling 
of the closed string metric, unlike the earlier case, we 
will now expand all results in powers of the scale 
parameter for closed string metric, $\ep$. We begin by expanding the DBI 
action,

\beqa
{\cal 
L}=\sqrt{(2\pi\al)e^{-(P+Q)\phi}(B+b)}\left[1+\f{1}{(2\pi\al)e^{-Q\phi}(B+b)}
\ep(\eta+h)\right]^{1/2}
\eeqa

For a matrix $M$, we have that following expansion,

\beqa
\sqrt{1+M}&=&\mbox{exp}\left[\f{1}{2}\Tr \mbox{log}(1+M)\right]\\
&=&1+\f{1}{2}\Tr(M-\f{M^2}{2}+...) 
+\f{1}{8}\left[\Tr(M-\f{M^2}{2}+...)\right]^2 + ...
\eeqa

For $M$  antisymmetric, terms containing $\Tr(M)$ vanishes, hence 
to order $\ep^2$, we have,

\beqa
{\cal L}&=&\sqrt{(2\pi\al)e^{-(P+Q)\phi}(B+b)}
\left[1-\f{\ep^2}{4}\Tr\left(\f{1}{(2\pi\al)e^{-Q\phi}(B+b)}(\eta+h)\right)^2
\right]\non
&=&\sqrt{(2\pi\al)e^{-(P+Q)\phi}(B+b)}\times\non
&\times&\left[1-\f{\ep^2e^{2Q\phi}}{4(2\pi\al)^2}
\Tr\left[\f{1}{B^2}\left(1-\f{2}{B}b+3\f{1}{B}b\f{1}{B}b-...\right)(\eta+h)^2
\right]\right]
\eeqa

Let us now first consider the ${\cal O}(1)$ term in $\ep$,

\beqa
{\cal L}\mid_{{\cal O}(1)}=\sqrt{(2\pi\al)e^{-(P+Q)\phi}(B+b)}
\eeqa

There is no graviton coupling at this order. The $\phi$ and $b$-field  
vertices from this are,

\beqa
V_{\phi}^1&=&-\f{1}{2}\sqrt{(2\pi\al)B}\left(\f{1}{B}\right)^{lk}\\
V_b^1&=&\sqrt{(2\pi\al)B}\left[\f{1}{4}\left(\f{1}{B}\right)^{lk}
\left(\f{1}{B}\right)^{ji}-\f{1}{2}\left(\f{1}{B}\right)^{jk}
\left(\f{1}{B}\right)^{li}
\right]
\eeqa

Now, let us consider the $\ep^2$ term. As in the earlier case let us 
define,

\beqa
Y&=&\sqrt{(2\pi\al)e^{-(P+Q)\phi}(B+b)}\\
X&=&-\f{\ep^2e^{2Q\phi}}{4(2\pi\al)^2}
\Tr\left[\f{1}{B^2}\left(1-\f{2}{B}b+3\f{1}{B}b\f{1}{B}b-...\right)(\eta+h)^2
\right]
\eeqa

We are interested in the two point function only upto ${\cal O}(\ep^2)$, 
hence we 
need not consider the graviton vertex. 
Also the $b$-field propagator has a $\ep^2$ factor (\ref{bp}). So, it 
is only 
necessary to compute the dilaton vertex at this order.
Listing the required derivatives,

\beqa
\f{\de Y}{\de \phi}=-\sqrt{(2\pi\al)B} \mbox{\hspace{0.2in}} \f{\de 
Y}{\de 
b_{kl}}=\f{1}{2}\sqrt{(2\pi\al)B}\left(\f{1}{B}\right)^{lk} 
\eeqa

\beqa
\f{\de^2Y}{\de b_{kl}\de \phi}=V_{\phi}^1 
\mbox{\hspace{0.2in}}
\f{\de^2X}{\de b_{kl} \de
\phi}=\f{\ep^24Q}{4(2\pi\al)^2}\left(\f{1}{B^3}\right)^{lk}
%\f{\de^2Y}{\de b_{kl}\de 
% b_{ij}}=V_{b}^1
\eeqa

\beqa
\f{\de X}{\de \phi}=-\f{\ep^22Q}{4(2\pi\al)^2}\Tr\f{1}{B^2}
\mbox{\hspace{0.2in}} \f{\de X}{\de 
b_{kl}}=\f{\ep^22}{4(2\pi\al)^2}\left(\f{1}{B^3}\right)^{lk}
\eeqa

After putting in all the appropriate derivatives, the 
vertices for the dilaton and the $b$-field upto ${\cal O}(\ep^2)$ is 
given by, 

\beqa
V_{\phi}&=&\sqrt{(2\pi\al)B}\left[-\f{1}{2}\left(\f{1}{B}\right)^{lk}
+\f{\ep^2(4Q-2)}{4(2\pi\al)^2}\left(\left(\f{1}{B^3}\right)^{lk}
-\f{1}{4}\Tr\left(\f{1}{B^2}\right)\left(\f{1}{B}\right)^{lk}\right)\right]\non
V_b&=&V_b^1
\eeqa

\begin{figure}[t]
\begin{center}
\epsfig{file=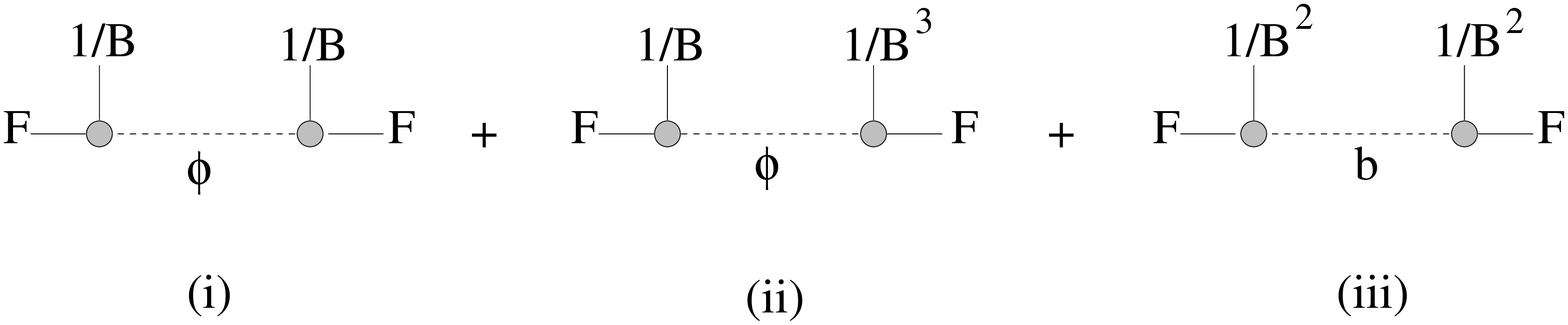, width= 12 cm,angle=0}
\vspace{ .1 in }
\begin{caption}
{Two point amplitude upto ${\cal O}(\ep^2)$. (i) and (ii) are due to
dilaton exchange, (iii) is due to $b$-field exchange.}
\end{caption}
\end{center}
\label{insertion}
\end{figure}

The situation in this case is similar to that of the earlier small $B$ 
expansion and is shown in Figure 2.
The propagators in this limit, eqns(\ref{dp2},\ref{bp}),

\beqa
<\phi\phi>&=&-\f{(D-2)i\ka^2}{4}\f{1}{\kpe^2+\ep^{-1}\kpa^2}\\
<b_{ij}b_{i^{'}j^{'}}>&=&-\f{2i\ka^2\ep^2}{(2\pi\al)^2}
\f{[\eta_{ii^{'}}\eta_{jj^{'}}-\eta_{ji^{'}}
\eta_{ij^{'}}]}{\kpe^2+\ep^{-1}\kpa^2}
\eeqa

With the vertices computed above and the propagator in this limit, 
the two point function for the dilaton is,

\beqa
L_2(b\phi 
b)&=&-i\mbox{det}(2\pi\al 
B)\ka^2\tau_p^2\f{(D-2)}{4}\int 
\f{d^l\kpe}{(2\pi)^l}\f{1}{\kpe^2+\ep^{-1}p^2} \times 
\non
&\times&\left[\f{1}{8}\left(\f{1}{B}\right)^{lk}\left(\f{1}{B}\right)^{l^{'}k^{'}}
-\f{\ep^2(4Q-2)}{8(2\pi\al)^2}\left(\left(\f{1}{B^3}\right)^{lk}
-\f{1}{4}\Tr\left(\f{1}{B^2}\right)\left(\f{1}{B}\right)^{lk}\right)
\left(\f{1}{B}\right)^{l^{'}k^{'}} \right]\non
&+& (lk) \leftrightarrow (l^{'}k^{'})
\eeqa

For the $b$-field,

\beqa
L_2(bbb)&=&-i\mbox{det}(2\pi\al 
B)\ka^2\tau_p^2\f{\ep^2}{(2\pi\al)^2}\int
\f{d^l\kpe}{(2\pi)^l}\f{2}{\kpe^2+\ep^{-1}p^2}
\times \non
&\times&[\left(\f{1}{4}\left(\f{1}{B^3}\right)^{lk}
-\f{1}{16}\Tr\left(\f{1}{B^2}\right)\left(\f{1}{B}\right)^{lk}\right)
\left(\f{1}{B}\right)^{l^{'}k^{'}} \non
&+&\f{1}{8}\left(\f{1}{B^2}\right)^{kk^{'}}\left(\f{1}{B^2}\right)^{ll^{'}}
-\f{1}{8}\left(\f{1}{B^2}\right)^{k^{'}l}\left(\f{1}{B^2}\right)^{lk^{'}}
\non
&+& (lk) \leftrightarrow (l^{'}k^{'})]
\eeqa

The first two terms cancel with the $Q$-dependent terms of the dilaton, 
the resulting amplitude can now be written as,

\beqal{final2}
L_2&=&-i\mbox{det}(2\pi\al B)\ka^2\tau_p^2
\int\f{d^l\kpe}{(2\pi)^l}\f{1}{\kpe^2+\ep^{-1}p^2}
[{\cal O}(1)+{\cal O}(\ep^2)]
\eeqa

\noindent
where,

\beqa
{\cal O}(1)=
\left[\f{(D-2)}{32}
\left(\f{1}{B}\right)^{lk}\left(\f{1}{B}\right)^{l^{'}k^{'}}
+(lk) \leftrightarrow (l^{'}k^{'})\right]
\eeqa

\beqa
{\cal O}(\ep^2)&=& 
\f{\ep^2}{(2\pi\al)^2}\f{(D-2)}{16}\left[\left[\left(\f{1}{B^3}\right)^{lk}
-\f{1}{4}\Tr\left(\f{1}{B^2}\right)\left(\f{1}{B}\right)^{lk}\right]
\left(\f{1}{B}\right)^{l^{'}k^{'}}\right]\non 
&+&\f{\ep^2}{(2\pi\al)^2}\left[
\f{1}{4}\left(\f{1}{B^2}\right)^{kk^{'}}\left(\f{1}{B^2}\right)^{ll^{'}}
-\f{1}{4}\left(\f{1}{B^2}\right)^{k^{'}l}\left(\f{1}{B^2}\right)^{lk^{'}}
\right]\non
&+& (lk) \leftrightarrow (l^{'}k^{'})
\eeqa

\noindent
We can now reconstruct the quadratic term in effective action, 
(\ref{eff}) following the earlier case.
With the following expansions, it is easy to check that the sum of the 
massless contributions adds upto eqn(\ref{clchannel}).

\beqa
G^{ij}&\sim &-\f{\ep}{(2\pi\al)^2}\left(\f{1}{B^2}\right)^{ij} +{\cal
O}(\ep^3)\\
\th^{ij}&\sim &
\left(\f{1}{B}\right)^{ij}+ 
\f{\ep^2}{(2\pi\al)^2}\left(\f{1}{B^3}\right)^{ij}\\
\sqrt{\ep\eta+(2\pi\al)B}&\sim &
\sqrt{(2\pi\al)B}\left[1-\f{\ep^2}{4(2\pi\al)^2}\Tr\left(\f{1}{B^2}\right)
\right]
\eeqa

\noindent
Note that, at the tree-level, to the linear order, $\hat{F}=F$, 
(\ref{redeff}). At this 
quadratic order in the effective action there is no need for 
redefinition of $F$ to equate the result here with that of string theory 
result in eqn(\ref{clchannel}).

\subsection{Noncommutative case ($G=\eta$)}

In this part we finally consider the restriction of the open string 
metric, $G=\eta$.
The lowest order solution for the closed string metric, $g$ in $\al$ in 
this limit is,

\beqal{soln}
g=-(2\pi\al)^2B^2 +{\cal O}(\alpha^{'4})
\eeqa

We will now consider expansions of the two point functions in 
powers of $\al$.
We begin again with the following DBI Lagrangian,

\beqa
{\cal L}&=&\sqrt{(2\pi\al)e^{-(P+Q)\phi}(B+b)}
\left[1-\f{1}{e^{-Q\phi}(B+b)}(2\pi\al)B^2(\eta+h)^2\right]^{1/2}
\eeqa

The calculation for the vertices is same as before, there is no graviton 
vertex to the leading orders. The dilaton and the $b$-field vertices 
are,

\beqa
V_{\phi}&=&\sqrt{(2\pi\al)B}\left[-\f{1}{2}\left(\f{1}{B}\right)^{lk}+
\f{(2\pi\al)^2(4Q-2)}{4}\left(B^{lk}-\f{1}{4}\Tr(B^2)\left(\f{1}{B}\right)^{lk}
\right)\right]\non
V_b&=&\sqrt{(2\pi\al)B}\left[\f{1}{4}\left(\f{1}{B}\right)^{lk}
\left(\f{1}{B}\right)^{ji}-\f{1}{2}\left(\f{1}{B}\right)^{jk}
\left(\f{1}{B}\right)^{li}
\right]
\eeqa

The propagators for the dilaton and the $b$-field are modified as,

\beqa
<\phi\phi>&=&-\f{(D-2)i\ka^2}{4}\f{1}{\kpe^2+\tilde{\kpa}^2/(2\pi\al)^2}\\
<b_{ij}b_{i^{'}j^{'}}>&=&-2i\ka^2(2\pi\al)^2
\f{[B^2_{ii^{'}}B^2_{jj^{'}}-B^2_{ji^{'}}
B^2_{ij^{'}}]}{\kpe^2+\tilde{\kpa}^2/(2\pi\al)^2}
\eeqa

\begin{figure}[t]
\begin{center}
\epsfig{file=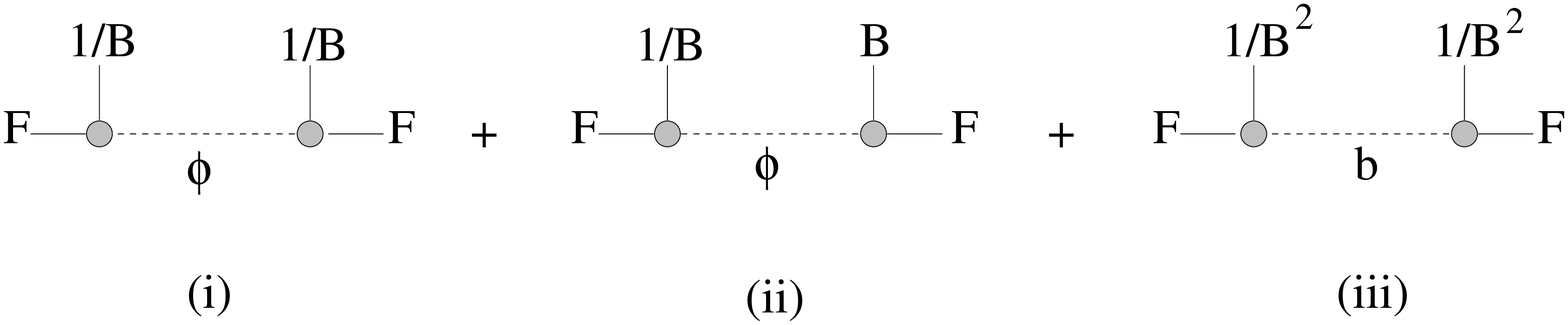, width= 12 cm,angle=0}
\vspace{ .1 in }
\begin{caption}
{Two point amplitude upto ${\cal O}(\als)$. (i) and (ii) are due to
dilaton exchange, (iii) is due to $b$-field exchange.}
\end{caption}
\end{center}
\label{insertion}
\end{figure}

With these vertices (shown in Figure 3) and the propagators from 
eqns(\ref{dp3},\ref{bp}), 
the two point functions are now given by,

\beqa
L_2(b\phi b)&=&-i\mbox{det}(2\pi\al 
B)\ka^2\tau_p^2\f{(D-2)}{4}\int \f{d^l\kpe}{(2\pi)^l}
\f{1}{\kpe^2+\tilde{p}^2/(2\pi\al)^2}\times\non &\times&
\left[\f{1}{8}\left(\f{1}{B}\right)^{lk} 
\left(\f{1}{B}\right)^{l^{'}k^{'}}-
\f{(2\pi\al)^2(4Q-2)}{8}\left(B^{lk}-\f{1}{4}\Tr(B^2)\left(\f{1}{B}\right)^{lk}
\right)\left(\f{1}{B}\right)^{l^{'}k^{'}}\right]\non
&+& (lk) \leftrightarrow (l^{'}k^{'})
\eeqa

\beqa
L_2(bbb)&=&-i\mbox{det}(2\pi\al B)\ka^2\tau_p^2(2\pi\al)^2
\int \f{d^l\kpe}{(2\pi)^l}
\f{2}{\kpe^2+\tilde{p}^2/(2\pi\al)^2}\times\non &\times&
\left[\left(\f{1}{4}B^{lk}-\f{1}{16}\Tr(B^2)\left(\f{1}{B}\right)^{lk}\right)
\left(\f{1}{B}\right)^{l^{'}k^{'}}
+\f{1}{8}(\eta^{ll^{'}}\eta^{kk^{'}}
-\eta^{kl^{'}}\eta^{lk^{'}})\right]\non &+& (lk) \leftrightarrow 
(l^{'}k^{'})
\eeqa

\noindent
As before, the first term of the $b$ exchange cancels with the $Q$ 
dependent term of the dilaton exchange. The full two point answer is

\beqal{final3}
L_2=-i\det(2\pi\al 
B)\ka^2\tau_p^2\int 
\f{d^l\kpe}{(2\pi)^l}\f{1}{\kpe^2+\tilde{p}^2(2\pi\al)^2}[{\cal 
O}(1)
+{\cal O}(\als)]
\eeqa

\beqa
{\cal O}(1)=
\left[\f{(D-2)}{32}
\left(\f{1}{B}\right)^{lk}\left(\f{1}{B}\right)^{l^{'}k^{'}}
+(lk) \leftrightarrow (l^{'}k^{'})\right] 
\eeqa

\beqa
{\cal O}(\als)&=&
(2\pi\al)^2\f{(D-2)}{16}\left[\left[B^{lk}
-\f{1}{4}\Tr(B^2)\left(\f{1}{B}\right)^{lk}\right]
\left(\f{1}{B}\right)^{l^{'}k^{'}}\right] \non
&+&(2\pi\al)^2\left[
\f{1}{4}\left(\eta^{ll^{'}}\eta^{kk^{'}}
-\eta^{kl^{'}}\eta^{lk^{'}}\right)
\right]\non
&+& (lk) \leftrightarrow (l^{'}k^{'})
\eeqa

\noindent
We will need the following expansions in this limit, to 
expand the closed string channel result upto this order.
We have already set,

\beqa
G^{ij}&=&\eta^{ij}
\eeqa

\noindent
and with the solution for $g$, eqn(\ref{soln}) to the lowest order in 
$\al$,

\beqa
\th^{ij}&\sim & \left(\f{1}{B}\right)^{ij}+(2\pi\al)^2B^{ij}\\
\sqrt{g+(2\pi\al)B}&\sim&
\sqrt{(2\pi\al)B}\left[1-\f{(2\pi\al)^2}{4}\Tr(B^2)\right]
\eeqa

\noindent
As in the earlier cases, the massless contributions computed here, 
eqn(\ref{final3}) adds upto eqn(\ref{clchannel}).
Note that the situation here is similar to that of the 
earlier case in section(4.2). As $\al \sim \sqrt{\ep}$, the 
closed string metric in both the cases goes to zero as $g\sim \ep$.
However the difference being that the two point amplitude differ
by powers of $B$ in both the cases, due to the  
relative power of $B^2$ in $g$ in this case. Here too, the SW 
map 
between the usual and the noncommutative field strength 
eqn(\ref{redeff}), remains the same. The differences in the powers of 
$B$ in the two point amplitudes, eqn(\ref{final2}) and 
eqn(\ref{final3}) are absorbed in $G$, $\theta$ and 
$\sqrt{g+(2\pi\al)B}$ in the two cases. We can work with any of the 
forms of the closed string metric $g$, the important point being that 
$g$ should go to zero as $\ep$ which gives the noncommutative gauge 
theory on the brane.

\section{Discussions}
\limits{}

\end{document}